\documentclass[%
 reprint,
 amsmath,amssymb,
 aps,
]{revtex4-2}
\pdfoutput=1 
\usepackage{graphicx}
\usepackage{dcolumn}
\usepackage{bm}
\usepackage{chemformula}
\usepackage{amsmath,scalerel}
\usepackage{braket}
\usepackage{subfigure}
\usepackage{cleveref}
\usepackage{flushend}
\usepackage{verbatim}
\usepackage{xr}
\usepackage[T1]{fontenc}
\usepackage[utf8]{inputenc}
\usepackage[normalem]{ulem}

\usepackage{changes}
\usepackage{cancel}
\usepackage{xcolor}
\makeatletter
\begin{document}

\preprint{APS/123-QED}

\title{
Guest-induced phase transition leads to polarization enhancement in MHyPbCl$_3$}

\author
{Pradhi Srivastava$^{1}$, Sayan Maity$^{2}$, Varadharajan Srinivasan$^{2}$}
\affiliation{$^{1}$Department of Physics, Indian Institute of Science Education and Research Bhopal, Bhopal 462 066, India}
\affiliation{$^{2}$Department of Chemistry, Indian Institute of Science Education and Research Bhopal, Bhopal 462 066, India}




\date{\today}

\begin{abstract}
	We investigate the structural and electronic changes in a 3D hybrid perovskite, methylhydrazinium lead chloride (\ch{MHyPbCl3}) from a \textit{host/guest} perspective as it transitions from a highly polar to less polar phase upon cooling, using first-principles calculations. The two phases vary structurally in the orientation of the \textit{guest} (MHy) and the two differently distorted \textit{host} (lead halide) layers. Our findings highlight the critical role of \textit{guest} reorientation in reducing \textit{host} distortion at high temperatures, making the former the primary order parameter for the transition, a notable contrast to the case of other hybrid perovskites. This is also confirmed by the dominating contribution of the \textit{guest} reorientation along the transition pathway. Analysis using maximally localized Wannier functions reveals that polarization enhancement upon heating is primarily due to the \textit{host} atoms, particularly of the more distorted octahedral layer. Despite its pivotal role in the transition, the contribution of \textit{guest} to polarization is relatively weaker, in contrast to previous suggestions. Furthermore, a significant ($\sim 9\%$) feedback polarization induced on the \textit{guest} by \textit{host} distortion significantly alters the density of states occupied by \textit{guest} closer to band-edges, suggesting a non-trivial contribution of \textit{guest} in impacting the optoelectronic properties and exciton binding energies.

\end{abstract}

\maketitle


\section{\label{sec:intro}Introduction}

Organic-inorganic halide perovskites (OIHPs) possess additional structural and chemical flexibility, compared to conventional metal halide perovskites~\cite{lee2012efficient,xu2019hybrid,petersson1988complete}. Indicated by chemical formula ABX$_3$, these materials are similar to \textit{guest-host} systems, where the rigid inorganic sub-lattice (B cation and X anion) \textit{hosts} organic \textit{guest}s (A cation) having translational and orientational degrees of freedom. \textit{Guest-host} systems are characterized by complexes involving two or more molecules or ions interacting with each other by forces other than covalent bonds, such as hydrogen bonds, van der Waals (vdW) interactions, or coordinate bonds~\cite{ryan2023quantifying,khan2021supramolecular}. This intricate interplay of forces is exemplified in hybrid perovskites, where a negatively charged inorganic framework acts as the \textit{host}, encapsulating positively charged organic molecules that act as the \textit{guest}s. The opposite polarities of \textit{guest} and \textit{host} ensure charge neutrality and overall stability of the crystal formed in a manner similar to that in their inorganic counterparts. However, in these hybrid structures, the stable orientations of the \textit{guest} molecules are significantly influenced by hydrogen bonding interactions with the \textit{host}. For instance, in the case of \ch{MA/FAPbX3}~\cite{lee2012efficient,xu2019hybrid,petersson1988complete}, these hydrogen bonds guide the \textit{guest} molecules into specific orientations within the \textit{host} framework. Moreover, as suggested by previous reports,~\cite{maity2024cooperative} these interactions can cooperate over large length scales, leading to spatially correlated \textit{guest} orientations and \textit{host} distortions. This cooperativity manifests globally as a structural phase of the system. Hence, hybrid perovskites could indeed be viewed from the point of view of \textit{guest-host} systems. While this perspective might be unconventional, it aligns well with established notions of \textit{guest-host} complexes.
Furthermore, such a perspective also serves as a conceptual framework wherein the response of the crystal to external perturbations can be rationalized in terms of the responses of the individual \textit{host/guest} components.  For instance, it is quite insightful to analyze the T-P phase diagram of a hybrid perovskite in terms of the expected phases of the individual sub-lattices. This would allow us to identify any emergent phenomena in the coupled system. Identification of key \textit{host} or \textit{guest} features influencing the stability of a phase has the practical advantage of offering strategies to design/tune \textit{guest}s and/or \textit{host}s, which can extend the stability of a phase or even suppress it completely. Experimentally, this would correspond to strategies such as doping, encoding organic \textit{guest}s with desired functionalities, applying strain, etc. This would, in turn, provide the flexibility to tune their electronic and optical properties according to different requirements. 

Along with electrostatic interactions, hydrogen bonds (H-bonds) and co-ordinate bonds between \textit{host} and \textit{guest} mediate their coupling resulting in various phases with variations in temperature.
%
%
%
Usually, stability of such phases capable of accommodating \textit{host} distortion is predicted by relative sizes of \textit{host/guest} ions indicated by Goldschmidt tolerance factor ($\tau$)~\cite{saparov2016organic}. In such stable OIHPs ($0.8\leq \tau\leq 1.0$), the primary order of phase transitions~\cite{cowley1980structural} is associated with \textit{host} sub-lattice (octahedral tilt modes)~\cite{mkaczka2022temperature,wu2023phase,saidi2016nature,hehlen2022pseudospin,li2023ferroelectric}. The orientational freedom of \textit{guest} plays a secondary role~\cite{parlinski1985secondary,cowley1980structural} in the overall phase transition through an order-disorder mechanism~\cite{maity2022deciphering}. \textit{Guest}-reorientation rate controlled by the activation barrier between two stable hydrogen-bonded orientations of the \textit{guest}, can be regulated  
by tuning the \textit{host} or \textit{vice-versa}~\cite{simenas2021phase,kubicki2017cation}. 
However, for a given \textit{guest}, the trend may be different depending upon several parameters, such as hydrogen bond strength, electronegativity, and polarizability of the \textit{host}~\cite{sharma2022influence,selig2017organic}. Alternatively, tuning the cation size, as shown mostly in mixed halide perovskites, can also modify the reorientation rate~\cite{mozur2021cation,mondal2024structural}. Hence, tuning the \textit{guest/host} chemistry can provide control over the driving force of structural phase transitions in OIHPs.

To understand the effect of \textit{guest} sizes, we classify OIHPs into normal-tolerant and over/super-tolerant perovskites, depending on whether $\tau$ remains within the conventional range or not. 
The phase transition temperature generally increases with the tolerance factor in a given series with varying \textit{guests}, which is indicative of stronger interaction between the \textit{guest} and the \textit{host} for higher $\tau$ (Tables~S1--S3). Usually, in normal-tolerant perovskites, significant \textit{host} transition barrier (defined here as the difference in energies of phases with distorted and undistorted \textit{host} lattice) implies their primary role in the phase transition mechanism. However, if the transition barrier of \textit{guest} is more than the \textit{host} barrier, \textit{guest} orientation can decide \textit{host} geometry, resulting in the reversal of the roles played by both sub-lattices. 
Strong coupling leads to locking of \textit{guest} orientations, usually seen in OIHPs with relaxed $\tau$ and decreased dimensionality owing to the presence of large \textit{guest} molecules ~\cite{fu2019incorporating}. This can also give rise to a polar/ferroelectric order, thereby facilitating electron-hole separation~\cite{liu2015ferroelectric, sherkar2016can}. Consequently, these materials are potential candidates for photovoltaics~\cite{chen2015ferroelectric,jena2019halide,green2014emergence,jeong2020stable,kim2022conformal}.

Such polar order in OIHPs is known to arise primarily due to the \textit{guest} dipoles through both translational and rotational modes~\cite{hu2017dipole}. They also cause polar distortions in \textit{host}-octahedra~\cite{huang2022understanding, srikanth2020optical,maity2023stabilizing} as a ``primary effect'', whose contribution to total polarization is usually small~\cite{frost2014atomistic}. In the limit of strong \textit{host/guest} coupling, a significant \textit{host} distortion can further affect the \textit{guest} polarization as a ``secondary effect'' which can be termed as ``feedback polarization'', that has so far not been reported. Thus, segregating the contributions of all these components is essential in order to tune the polarization.  

\begin{figure*}
	\centering
	\hspace{0cm}
	\subfigure[Phase-I ($(I)_{11}$)]{
		\includegraphics[width=0.70\columnwidth, keepaspectratio]{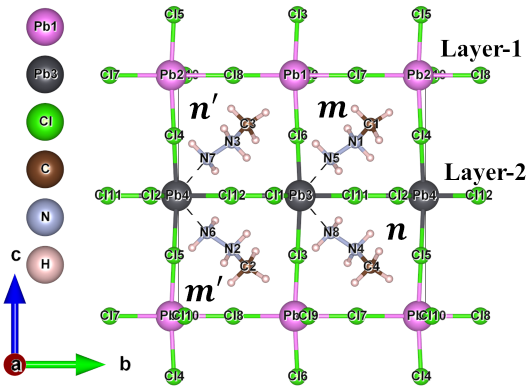}
		\label{subfig:ht_opt}
	}
	\hspace{0cm}
	\subfigure[Phase-II ($(II)_{21}$)]{
		\includegraphics[width=0.5\columnwidth, keepaspectratio]{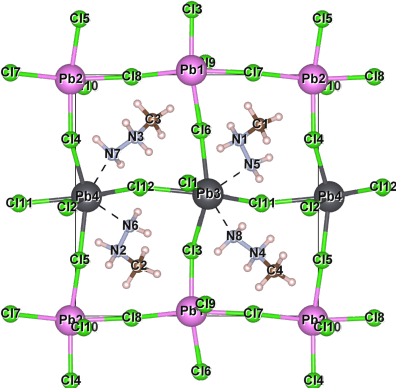}
		\label{subfig:rt_opt_transf}
	}
	\hspace{0cm}
	\subfigure[$(\sim II)_{21}$]{
		\includegraphics[width=0.5\columnwidth]{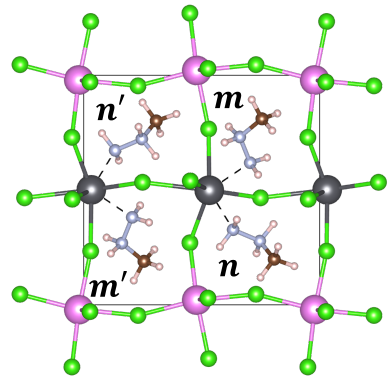}
		\label{subfig:12_opt_numbered}
	}
	\caption{Fully optimized structures of \ch{MHyPbCl3} for both the experimental phases. Pink \ch{Pb} atoms correspond to the less distorted octahedral layer (Layer-1) and grey \ch{Pb} atoms to the more distorted octahedral layer (Layer-2).~\subref{subfig:12_opt_numbered} Phase-II optimized under the constraints of Phase-I lattice parameters.}
	\label{fig:opt_ht_LT}
\end{figure*}  

Recently, a new series of  3D-OIHP \ch{MHyPbX3} (\ch{MHy}=methylhydrazinium, \ch{X=Br/Cl}) has been synthesized~\cite{maczka2020methylhydrazinium,maczka2020three,maczka2020nmr,maczka2022comparative,drozdowski2022three}, despite leading to breakdown of conventional tolerance factor ($\tau$) criteria 
due to the large ionic radius of the \textit{guest}. 
\ch{MHyPbCl3} undergoes an order-to-order transition from a low temperature (LT) less polar monoclinic ($P2_1$, Phase-II) to a high temperature (HT) more polar orthorhombic phase ($Pb2_1m$, Phase-I) at $\sim$342 K, thus showing polarity at room-temperature. The absence of proper ferroelectric behavior was demonstrated through the fact that no switching of polarization was observed on switching of external electric field. The closed polarization-electric field (P-E) loop originates from the conductivity of the sample. X-ray diffraction data also consistent with this observation since it shows that the transition from $Pb2_1m$ to $P2_1m$ is a $mm2F2$ type transition, which is not a ferroelectric one. Therefore, the inability of reversible switching of \ch{MHy+} cation upon application of an external field suggests that the \ch{MHy+} hybrid systems may exhibit improper ferroelectric behavior. 
Interestingly, \ch{MHyPbCl3} decomposes ($\sim490$ K) before a transition to a non-polar disordered state, unlike other conventional OIHPs or even \ch{MHyPbBr3}~\cite{maczka2020methylhydrazinium,maczka2020three}. The individual roles of \textit{host/guest} and their coupling in stabilizing the polar phases are as yet unclear. Particularly, the presence of a large \textit{guest} can have a significant role in the phase transition mechanism through modification of the rotational energy barrier.

Following these crucial observations from the experiments, we employ first-principles density functional theoretic (DFT) calculations on \ch{MHyPbCl3} to understand the primary order parameter responsible for driving the structural phase transition. We find that, unlike prototypical analogs, a relatively high \textit{guest}-reorientation barrier supports the primary role of the \textit{guest} in driving the transition in this strongly coupled system. We confirm this result by quantifying the relative extents of \textit{host} and \textit{guest} participation along the minimum energy path connecting the two phases.
Secondly, in contrast to the previous claims~\cite{maczka2020three}, we find that the \textit{host} is primarily responsible for the polarization enhancement with temperature making the system an improper ferroelectric.~\cite{maczka2020three} 
Using maximally localized Wannier functions (MLWFs) we not only unequivocally establish the dominant contribution of the \textit{host} to the polariztion enhancement 
but also show that the \textit{host} distortion accompanying the transition induces a positive polarization feedback of $\sim9\%$ on the \textit{guest} ions. This feedback manifests itself through modulation of \textit{guest} states at the band edges, which are usually dominated by the \textit{host} contributions. 
Such a mechanism allows the direct influence of the organic \textit{guest} in tuning the electronic and optical properties of over-tolerant OIHPs, making them appealing for optoelectronic applications.

\section{Results and discussion}
\subsection{\label{sec:results_discussion}Structural optimization}

The structures of both Phase-I and Phase-II of \ch{MHyPbCl3} were fully relaxed at $0$ K, starting from the respective experimental structures. The van der Waals correction is accounted for by including Grimmes D2 functional~\cite{grimme2006semiempirical,barone2009role}. The lattice parameters thus obtained match well with experimental values within accepted tolerance of $\pm5~\%$~\cite{zhang2018phase,da2015phase} irrespective of the choice of exchange-correlation (XC) functional (Table~S4, also see~Eq.~S.2, Table~S5 and Table~S6). However, we chose to work with PBEsol~\cite{perdew2008restoring} as it not only predicts the relative order of lattice constants ($a<b<c$ for Phase-II) accurately, but also has been preferred in other halide perovskites~\cite{hernandez2019dft,jong2019first,castelli2014bandgap,agbaoye2021band}. 
%


The optimized structures (Fig.~\ref{fig:opt_ht_LT}) of the two phases differ in both the \textit{guest} (\ch{MHy+}) orientation and the \textit{host} (\ch{PbCl6} octahedra) distortion, confirming orthorhombic structure ($Pb2_1m$) for Phase-I and monoclinic ($P2_1$) for Phase-II, similar to experiment~\cite{maczka2020three}. Phase-II is the ground state at $0K$, being lower in energy than Phase-I by $\sim 45~$meV/f.u. 
%
%

\begin{figure}
	\centering
	\includegraphics[width=0.4\textwidth]{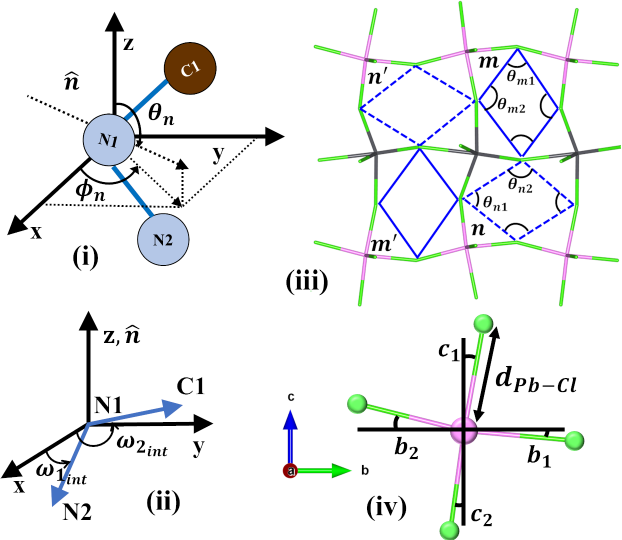} 
	
	\caption{\label{subfig:oct_tilt_schematic}~(i)-(ii) Parameters to describe the \textit{guest} orientation, (iii) \textit{host}/octahedral tilting angle, $\delta \theta_{tilt}$, measured as the average deviation from 90 of the internal angles $\theta$, and (iv) alternate measure for tilting angle in terms of Glazer angles given by $\delta \theta_{gl}=(|b_1|+|b_2|+|c_1|+|c_2|)/4$ and \textit{host} distortion indicated by the rms deviation of bond lengths, $\delta d_{\mathrm{Pb-Cl}}$.}
\end{figure}

We define a few parameters to distinguish the \textit{host/guest} configurations between different phases, as shown schematically in Fig.~\ref{subfig:oct_tilt_schematic}. We assign Glazer notation of $a^0b^-c^-$  to Phase-I since the \textit{host} tilting about $a-$axis is very small (nearly 0) compared to ideal perovskite (as shown in Fig.~S2(c) and Fig.S2(d)). Similarly, Phase-II has a Glazer notation of $a^+b^-c^-$. Hence, the octahedral tilting transition $a^0b^-c^-\to a^+b^-c^-$ is seen from Phase-I to Phase-II. However, two alternate layers of octahedra are identified along crystal $c$-axis in both phases due to coordinate bonds between the \textit{guest} and the \textit{host}, layer-1 being less distorted (pink atoms in Fig.~\ref{fig:opt_ht_LT} and pink octahedra in Fig.~S4) and layer-2 being more distorted (grey \ch{Pb} atoms and grey octahedra)~\cite{maczka2020three}. The difference in the two layers is shown by the variation in the bond lengths, $d_{\mathrm{Pb-Cl}}$ and \textit{host} tilting angles, $\delta \theta_{tilt}$ and $\delta \theta_{gl}$ (defined in Eqs.~S.4--S.6 and shown in Fig.~\ref{subfig:oct_tilt_schematic} and Fig.~S2) for different structures under consideration. In Phase-II (Fig.~\ref{subfig:rt_opt_transf}), half of the \ch{MHy} sitting alternately along $b$ and $c$ axes (or at $m-m'$), gets significantly reoriented from Phase-I, whereas those at $n-n'$ are similar to Phase-I. The \textit{host} and the \textit{guest} configurations have been defined in Sec.~S-II and have been tabulated for different structures (Table~S10).
%
%
%
%

%
As any phase can differ by at least one of the \textit{host/guest} sub-lattices, we employ the following nomenclature for identifying the structural differences in those components:
%
%
\[\scaleto{(Host)_{Guest(m-m'), \textit{Guest}(n-n')}}{20pt}\]
Here, the first symbol (Roman numeral) denotes the \textit{host} tilting ($\delta \theta_{tilt}$ close to either Phase-I or Phase-II), whereas the numbers in the subscript correspond to the \textit{guest} orientation at $m-m'$ and $n-n'$ sites, respectively (Fig.~\ref{subfig:ht_opt}). Structures with \textit{guest} orientation within $10^{\circ}$ of either phase are classified as being in that phase. Thus, Phase-I and Phase-II are indicated by $(I)_{11}$ and $(II)_{21}$, respectively.  Additionally, the `$\sim$' symbol before the phase denotes a constraint on the \textit{host} lattice parameters. For example, $(\sim II)_{21}$ corresponds to a structure with Phase-I lattice parameters, but the \textit{host} distortions and \textit{guest} orientations are similar to Phase-II. Such an optimization was deemed necessary to compare the structural, electronic, and polar properties of the two phases at the same volume. The relevant parameters are defined in Fig.~\ref{subfig:oct_tilt_schematic} and have been tabulated for different structures in Table~S10.

\subsection{Structural phase transition}
\subsubsection{\label{ssec:cor_oct_mhy} \textit{Host/Guest} coupling}
Electronic properties, such as the band gaps, in OIHPs are governed significantly by the \textit{host} atoms because of their high density of states near the band-edges~\cite{fan2015perovskites}. The band gaps can, thus, be modulated by having a control over the \textit{host} distortion which, in turn, is affected by the \textit{guest/host} coupling. In other analogs, it has been found that the \textit{guest} and \textit{host} act cooperatively,~\cite{li2016atomic} referred to as the `chicken (\textit{guest})-and-egg (\textit{host}) paradox of the organic-inorganic interaction'. We expect that in \ch{MHyPbCl_3}, being a super-tolerant stable perovskite, the HG coupling should be quite strong (which has been quantified in Sec.~SIII A, and compared across different \textit{guest}s in Table~S8). 
Hence, the impact of \textit{guest} on the \textit{host} should be more significant in comparison to other analogs. Therefore, understanding the phase transition mechanism in such a system would be quite interesting.

To this end, we performed a series of structural relaxations starting from different geometries and imposing constraints as required. As indicated above, full structural optimizations starting from experimental structures of HT and LT phases resulted in the $(I)_{11}$ and $(II)_{21}$ structures, respectively, which retained the corresponding experimental configurations of the \textit{host} and \text{guest}. We note here that the optimized $(I)_{11}$ was nevertheless found to be dynamically unstable, a point we will explore in more detail later. Interestingly, when the \textit{guest} configurations of the two phases were interchanged, the modified HT structure ($(I)_{21}$) yielded the LT structure upon full optimization.
In contrast, upon optimization,  $(II)_{11}$ retained nearly the same \textit{host} and \textit{guest} configurations, without any significant \textit{guest} reorientation irrespective of whether lattice parameters were allowed to relax (see Fig.~S3 and Fig.~S4, and Table~S11). 
However, starting an optimization from $(I)_{21}$ at fixed lattice parameters yielded $(\sim II)_{21}$ , a structure with Phase-II-like \textit{host} distortions as well as \textit{guest} orientations (shown in Fig.~\ref{subfig:12_opt_numbered}). Together with the fact that optimization starting from $(I)_{11}$ yielded the Phase-I structure, this implies that \textit{host} distortions are not energetically favoured without the \textit{guest} reorienting to Phase-II configuration. Hence, the \textit{guest} reorientation acts as a driving force for the structural transition, making it the primary order parameter. 
The interplay between \textit{guest} and the \textit{host} for this system is contrary to what has been reported for perovskites with smaller tolerance factors, (\ch{A}=\ch{MA+}, \ch{FA+}; \ch{B}=\ch{Pb}, \ch{Sn}, etc.) ~\cite{kim2020first,maity2022deciphering} because of the inability of \ch{MHy+} to rotate freely in the cuboctahedral void of the \textit{host}. This has also been confirmed later in this work by the presence of relatively high rotational energy barriers for the switching between the two phases. A one-to-one correspondence between the \textit{guest} orientation and the \textit{host} distortion, as shown in Fig.~S6 and Fig.~S7 suggest the involvement of a complex order parameter, which could involve significant contributions from both the \textit{guest} reorientation (predominantly) as well as the \textit{host} distortion indicating a strong \textit{guest/host} coupling. 

\begin{table*}
	\caption{\label{tab:rel_energy_guest_op}%
		Relative energies of final structures obtained by optimization starting from different initial configurations to establish the primary role of \textit{guest} in driving the structural phase transition. The optimization in the last row was performed, keeping the lattice parameters fixed at their values in Phase-I.}
	\centering
	\begin{tabular}{c c c}
		\hline
		\hline
		{\bf Initial}& {\bf Final} & $\mathbf{\Delta E=E-E_{(II)_{21}}}$ \\
		{\bf Configuration} & {\bf Configuration} & ( meV/unit )  \\
		\hline
		$(I)_{21}/(II)_{21}$&$(II)_{21}$&0.0 \\
		$(I)_{11}$&$(I)_{11}$&176.0 \\
		$(II)_{11}$&$(II)_{11}$&117.0 \\ 
		$(I)_{21}$&$(\sim II)_{21}$&143.0 \\
		\hline
	\end{tabular}
\end{table*}



A further confirmation regarding the primary role of the \textit{guest} reorientations in driving the transition is obtained
by quantifying the \textit{guest} reorientational barrier encountered during transition and the relative contribution of the two sub-lattices along the minimum energy path (MEP) of transition. For this and other investigations described below, we compare $(I)_{11}$ and the constrained Phase-II, $(\sim II)_{21}$, obtained by the optimization of $(I)_{11}$ at fixed lattice parameters (Fig.~\ref{subfig:12_opt_numbered}). This becomes specifically important for polarization calculations since the polarization quantum depends on the lattice parameters~\cite{spaldin2012beginner}. 

\subsubsection{\label{ssec:struc_transf}\textit{Guest}-reorientational energy barrier and Transition Character}
To estimate the rotational energy barrier of the \textit{guest}, the \textit{guest} is transformed between its orientations in the two phases using a combination of rotational and translational coordinates (R and T matrices). The details of the transformation are mentioned in Sec.~S-IV A. This can be used to disentangle the \textit{host} and \textit{guest} sub-lattices, as was done to separate the contributions of \ch{MA+} and inorganic \ch{BX3} to polarization~\cite{stroppa2015ferroelectric}.

\begin{figure*}
	\centering
	\includegraphics[width=1.0\textwidth]{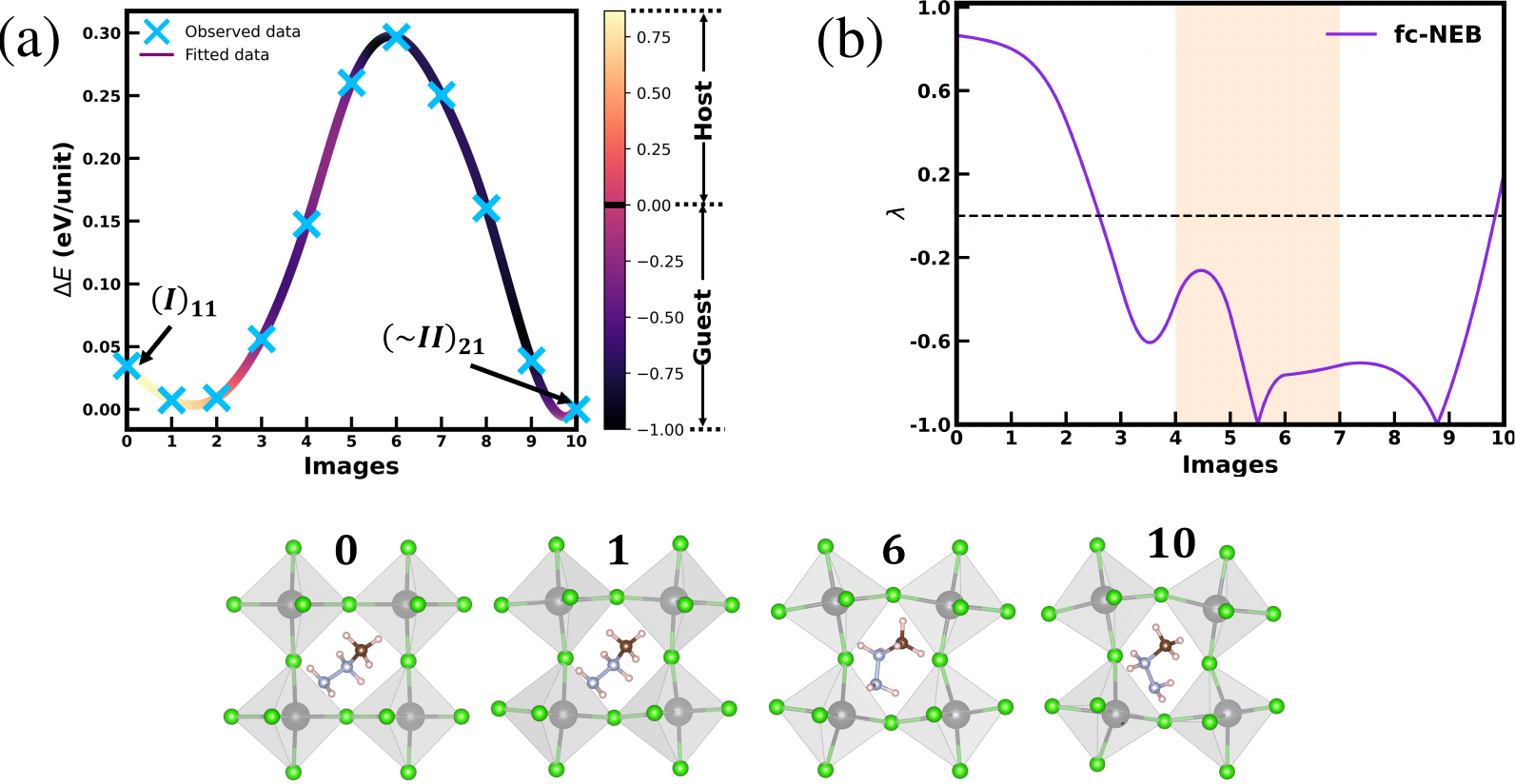}
	
	\caption{\label{fig:gh_param_variation}~(a)~Energy barrier (fitted by spline interpolation) along minimum energy path (MEP) obtained from fc-NEB (fixed cell NEB) path in Phase-I lattice parameters), the color code indicates the primary involvement of either \textit{guest} or \textit{host} in different regions of the curve. The structures shown below are the initial (Phase-I), $1^{st}$ (where \textit{host} distortion is prominent), $6^{th}$ (the transition point, $guest$ reorientation is dominant) and the final images $(\sim II)_{21}$. They indicate some of the crucial points during the transition.(b)~Transition character (TC), denoted by $\lambda$, as a function of the reaction coordinates along fc-NEB path. The values are derived from the interpolated data on $\theta_G$ and $\theta_H$. The shaded region marks the region around the transition point which is \textit{guest} dominated. The black dotted line (at $\lambda=0$) demarcates the \textit{host} and \textit{guest} dominated regions.}
\end{figure*}


Fig.~\ref{fig:gh_param_variation}(a) estimates the rotational energy barrier due to the orientational pattern of the \textit{guest} in $(I)_{11}$ and $~(\sim II)_{21}$ which is around $0.3$ eV/unit. This is obtained through Climbing Image-Nudged Elastic Band (CI-NEB) along the path determined by the rotation of the \textit{guest} in fixed cell parameters of Phase-I. This path is referred to as fixed-cell NEB (fc-NEB). This high value of the barrier prevents the reorientation of \textit{guest} at $0$ K irrespective of the starting structure. This also explains why the high temperature phase, Phase-I, remains intact even after cell optimization.
The variation in the \textit{host} tilting pattern and the \textit{guest} orientation has been tabulated in Table~S9 for the path obtained by fc-NEB. The \textit{host} variation (plotted in Fig.~S11(b)) shows that the \textit{host} stabilizes in its Phase-II-like octahedral pattern around the transition state (which is decided by the \textit{guest} orientation), beyond which the optimization of any of the images would eventually lead to Phase-II. \textit{Host} relaxation introduces a reduction in the rotational energy barrier, thereby facilitating its reorientation (Fig.~S11(a)).
To eliminate any effects due to the constraining of lattice parameters, the rotational energy barrier is also obtained by performing a variable-cell NEB from Phase-I to Phase-II (shown in Fig.~S12). The energy barrier between the two orderings of the \textit{guest} still is around $0.4$ eV/unit which further confirms the inability of the molecule to reorient itself significantly irrespective of its starting configuration and the lattice parameters. 

The driving force leading to structural phase transition can be quantified in terms of the relative contribution of the \textit{guest} and \textit{host} along MEP. For this, we define a parameter, Transition Character (TC), which is evaluated by the changes in the \textit{guest} or \textit{host} configuration as the transition progresses along the MEP. For the $i^{th}$ image we define the \textit{host} and \textit{guest} parameters $\Delta \theta_{H,i}$ 
and $\Delta \theta_{G,i}$
, respectively. Here, $\Delta \theta_{H,i}$ measures the differential change in $\delta \theta_{tilt}$ (see Fig.~\ref{eq:order_param}) normalized by total change in it across the transition. $\Delta \theta_{G,i}$ is calculated as
\begin{subequations}
	\begin{align}
	\Delta \theta_{G,i}&=\sqrt{(\Delta \omega^2_{norm,i}+\Delta \omega^2_{int,i})}
	\label{subeq:guest_dev1} 
	\end{align}
\end{subequations}
where $\Delta \omega_{norm,i}$ and $\Delta \omega_{int,i}$ are the differential changes in the molecular plane-normal and the internal angles of the \textit{guest}, respectively, normalized by their overall change across the transition (see Sec. SIV B for details).

We now define a quantity called the transition character (TC), denoted by $\lambda$, as the difference between the rates of configurational changes of \textit{host} and \textit{guest} to the sum of their absolute values, as given by Eq.~\ref{eq:order_param}.

\begin{equation}
\lambda_i=\frac{\left(|\Delta \theta_{H,i}| -|\Delta \theta_{G,i}|\right)}{(|\Delta \theta_{H,i}| +|\Delta \theta_{G,i}|)},~-1\leq \lambda_i \leq 1
\label{eq:order_param}
\end{equation}

\textit{Guest} orientation predominates over \textit{host} tilting for the regions where $-1<\lambda<0$, and \textit{host} distortion is dominant for $\lambda>0$, with both contributing equally at $\lambda=0$. 

Fig.~\ref{fig:gh_param_variation}(a) shows the energy barrier from Phase-I to $(\sim II)_{21}$ color-coded according to TC (the variation in TC is shown in Fig.~\ref{fig:gh_param_variation}(b) and Fig.~S13(c)).  The huge barrier seen is clearly associated with the \textit{guest} rotational energy cost. Hence the transition is mostly governed by \textit{guest}, thereby making it the primary order parameter of structural phase transition. The majority of the \textit{host} transition occurs near the first image, involving a soft mode arising from the instability in $(I)_{11}$ earlier and as discussed below. vc-NEB from Phase-I to Phase-II structure also confirms the primary involvement of \textit{guest} around the transition point (Fig.~S12, Fig.~S13(b) and Fig.~S13(c)). Additionally, in vc-NEB, the \textit{host} involvement also becomes prominent along the transition path in comparison to fc-NEB. However, in either of the cases, transition is not achievable until the molecule rotates by a critical angle, i.\,e., \textit{host} distortion alone cannot drive this phase transition.

\subsubsection{\textit{Host}-distortion}
\begin{figure}
	\centering
	\subfigure[Unstable phonon eigen mode of Phase-I]{
		\includegraphics[width=0.4\textwidth]{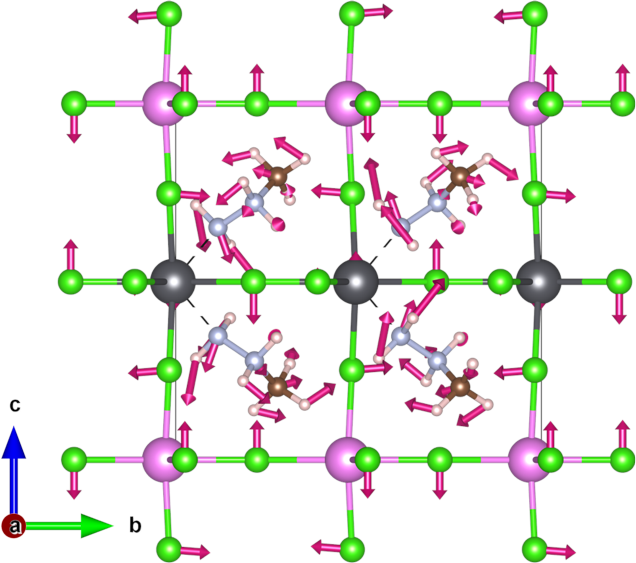}
		\label{spsubfig:ht_unstable_forces}} 
	\hspace{0cm}
	\subfigure[PES of unstable phonon mode. The dotted lines are a guide to the eye.]{\includegraphics[width=0.45\textwidth]{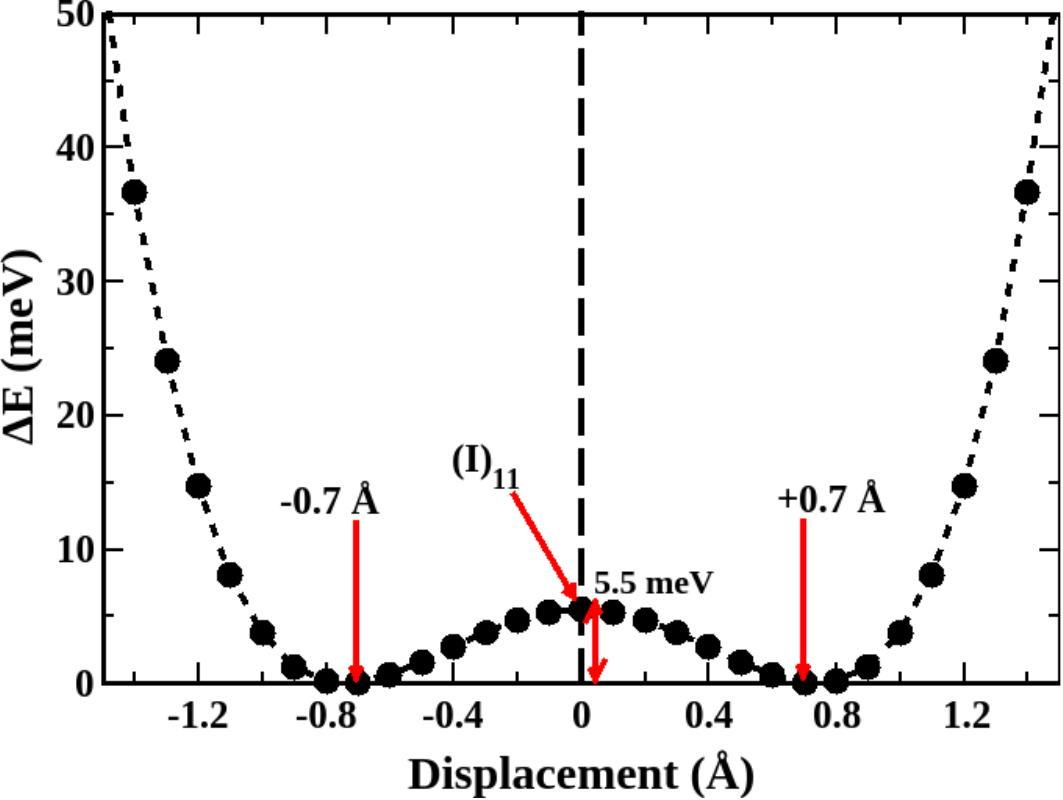}
		\label{subfig:unstable_mode_I}}
	\caption{\label{fig:unstable_mode_I}~Unstable phonon mode and PES of Phase-I. Magenta arrows in ~\subref{spsubfig:ht_unstable_forces} show the magnitude of forces on each of the atoms in the unstable mode. PES of the unstable phonon mode in ~\subref{subfig:unstable_mode_I} show two new energy-minimum states. The transition of \textit{host} of Phase-I to Phase-II occurs \textit{via} this mode.}
\end{figure}

To analyze the interplay between \textit{guest} and \textit{host}, we need to understand the involvement of \textit{host} as well. Therefore, we investigate the phonon dispersions of different structures of \ch{MHyPbCl3} considered here. We find that all structures 
are dynamically stable within typical numerical accuracy ($\sim0.3$ THz)~\cite{huan2016layered,marronnier2017structural} except for $(I)_{11}$ (Fig.~S14 and Fig.~S15). The latter displays an imaginary optical mode at $\Gamma$ (Fig.~S15(b)) originating on the \textit{host} as is expected by a similar origin of low-frequency modes in other halide perovskites~\cite{maczka2020methylhydrazinium,maczka2020nmr,marronnier2017structural,yaffe2017local}. 
Starting from this unstable Phase-I structure, each ion $i$ is displaced by $\textbf{u}_i (\eta)$ along the unstable optical mode, as is given by Eq.~\ref{speq:phonon_disp}. Here $x$ is the soft or the unstable mode eigenvector at $\Gamma$ and $\eta$ is the unitless displacement parameter~\cite{wang2021first}. 
\begin{equation}
\textbf{u}_i(\eta)=\eta \cdot \textbf{x}_{i\nu_{\Gamma}} (\textbf{q}_{\Gamma})
\label{speq:phonon_disp}
\end{equation}
The PES calculated along the soft/unstable eigenmode (Fig.~\ref{fig:unstable_mode_I}) shows a local maximum at Phase-I with minima at $\eta=\pm~0.7\text{\AA}$, corresponding to a maximum atomic displacement of around $0.1~\text{\AA}$. A very small energy barrier of $\approx1.4$ meV/f.u. (which reduces to $\approx0.36$ meV/f.u. with hybrid-functionals (Fig.~S15(b)))
suggests that Phase-I results from dynamic disordering of the \textit{host} over the two off-centred minima.

The structures of the two new energy equivalent minima are related by $b$-glide and 
are dynamically stable (Fig.~S15(c)). They have similar distortion pattern as Phase-II (Fig.~S16 and Fig.~S17) without significant reorientation of the \textit{guest}, hence are labeled as $(\sim II)_{11}$. In either structure, the
\textit{host} picks up the tilting pattern of Phase-II, indicating that Phase-I should be characterized by permanent \textit{host} distortions.
However, the reorientation of \textit{guest} remains hindered because of the huge rotational energy barrier as is discussed earlier. This further is an evidence of the dominant role of the \textit{guest} in driving the phase transition and confirmed by TC values along the transition path. The dynamical stability of the $(II)_{11}$ structure further confirms that the preferred state of the \textit{host} is the ordered state, irrespective of the lattice parameters (Fig.~S15(d)). Therefore, the experimental \textit{host} must be dynamically disordered over the two stable \textit{host}-distorted states, appearing as the experimental Phase-I structure on average.  

The independent analysis of \textit{guest} orientation as well as \textit{host} distortion provides us an insight into the significant roles of each of them in phase transition. Fig.~S18 summarizes the energy scales of all the structures that have been explored. 
The polar distortions can be induced in the \textit{host} at finite temperatures by sufficiently large electric fields~\cite{leppert2016electric}. Likewise, it can be anticipated that applying an electric field in a suitable direction will induce an asymmetry in the PES of Phase-I of \ch{MHyPbCl3}, leading to the preferential stabilization of one structure over another. Consequently, the control over the stabilization of the disordered \textit{host} within one of the ordered energy-minimum structures can be governed by the electric field. This can further lead to the development of extended hybrid networks with promising applications in ferroic-related fields.
\subsection{Polarization}
\begin{figure}
	\centering
	\includegraphics[width=0.48\textwidth]{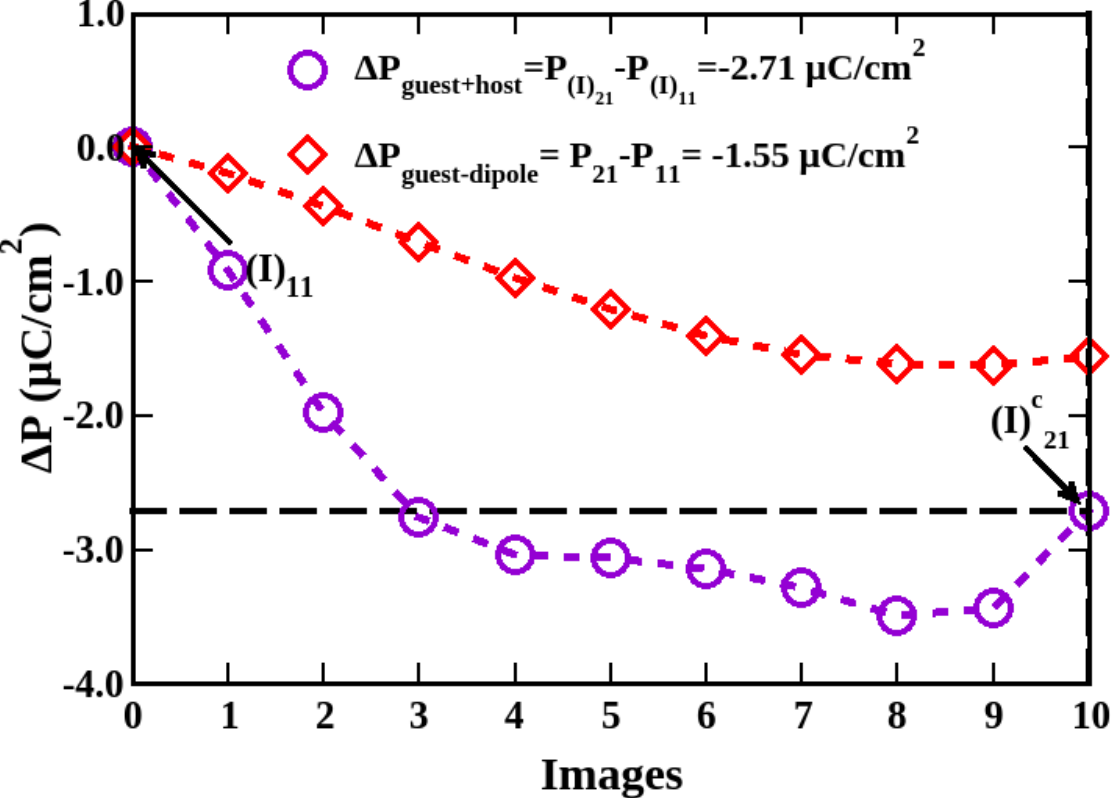}
	\caption{Polarization difference ($\Delta P$) for different images obtained by reorienting the \textit{guest} at $m-m'$. This transformation takes the structure $(I)_{11}\to(I)^c_{21}$, where `c' indicates that the \textit{guest} at $n-n'$ and \textit{host} are constrained in Phase-I, while the \textit{guest} at $m-m'$ are in $(\sim II)_{21}$. The red curve (diamond symbols) is for isolated dipoles placed in a positive background; the violet (circular symbols) is for the \textit{guest} in the presence of Phase-I \textit{host}. The dotted lines are a guide to the eye.}
	\label{fig:pol_mhy_berry_gaussian}
\end{figure}

Another interesting feature in \ch{MHyPbCl3} is the enhancement of polarization with the increase in temperature. \textit{Guest}-driven transitions can indirectly induce polarization in the \textit{host} through distortion. Here we probe into the origin and variation of polarization across different phases of this material to disentangle the \textit{guest/host} contribution. Tuning these structural features may lead to enhancement of polarization in OIHPs, thus improving PCEs~\cite{stroppa2015ferroelectric,huang2017understanding,shahrokhi2020emergence}. Here, we have employed Berry phase method~\cite{king1993theory,resta1994modern,resta2007theory,spaldin2012beginner} as well as maximally localized Wannier functions (MLWF)~\cite{marzari1997maximally,pizzi2020wannier90,mostofi2014updated} for the in-depth analysis of the polar order of \ch{MHyPbCl3}. The polarization values are converged with respect to the number of $k$-points (Fig.~S19). The values obtained from these two methods are in excellent agreement as reported in SI (see Table~S12 and Table~S13). The usage of the same lattice parameters (here, in Phase-I) ensures the same polarization quanta for different structural phases.
Since the phase transition is \textit{guest}-driven, we, first look at the effect on total polarization solely due to \textit{guest} reorientation (at $m-m'$), by evaluating the net dipole moment/polarization in the absence (isolated dipoles, Fig.~S20) and the presence of a fixed \textit{host} (here, Phase-I). The isolated dipoles show a huge variation in the dipole moment (shown by diamond symbols in Fig.~\ref{fig:pol_mhy_berry_gaussian}) and have the capability to induce large polarization. However, this is suppressed as soon as the \textit{host} is introduced \textit{via} polarization feedback to the \textit{guest} (as discussed later). The effect of aforementioned \textit{guest} reorientation in Phase-I \textit{host} is captured through a pathway from $(I)_{11}$ to $(I)^{c}_{21}$, as shown in Fig.~\ref{fig:pol_mhy_berry_gaussian}. Here, `c' indicates that the \textit{guest} at $n-n'$ and \textit{host} are constrained in Phase-I, while the \textit{guest} at $m-m'$ are in $(\sim II)_{21}$. This shows a net decrease in the total polarization of the combined \textit{guest/host} system which could be due to three reasons: (1) Phase-II-like reorientation of isolated dipoles (red curve, $\Delta P=1.55~\mu C/cm^2$), (2) Dipole-dipole interaction among the \textit{guest}s ($\Delta P=0.12~\mu C/cm^2$) and (3) \textit{Host} involvement, \textit{i.e.} the electronic charge density around the \textit{host} atoms induced by (1) ($\Delta P=1.16~\mu C/cm^2$). These contributions are independent of whether the \textit{guest} at $m-m'$ or $n-n'$ is rotated as seen from the symmetric dependence of $\Delta P$ in Fig.~S21. Introducing \textit{host} distortion in $(I)^c_{21}$ and optimization of the \textit{guest} at $n-n'$ leads to the final structure, $(\sim II)_{21}$, which has an effective polarization of $-2.57~\mu C/cm^2$.


For a series of similar OIHPs, a polarization range of 1-40~$\mu C/cm^2$ has been reported and these values hugely depend on the orientation of the local dipoles~\cite{stroppa2015ferroelectric,hu2017dipole}. We find that $\ch{MHyPbCl3}$ lies well within this range, thus having the potential for modulating/enhancing optoelectronic properties at room-temperature, unlike other reported OIHPs.

\subsubsection{\label{sssec:mlwf_pol}Disentangling different components of polarization}

\begin{figure}
	\centering
	\includegraphics[width=0.45\textwidth]{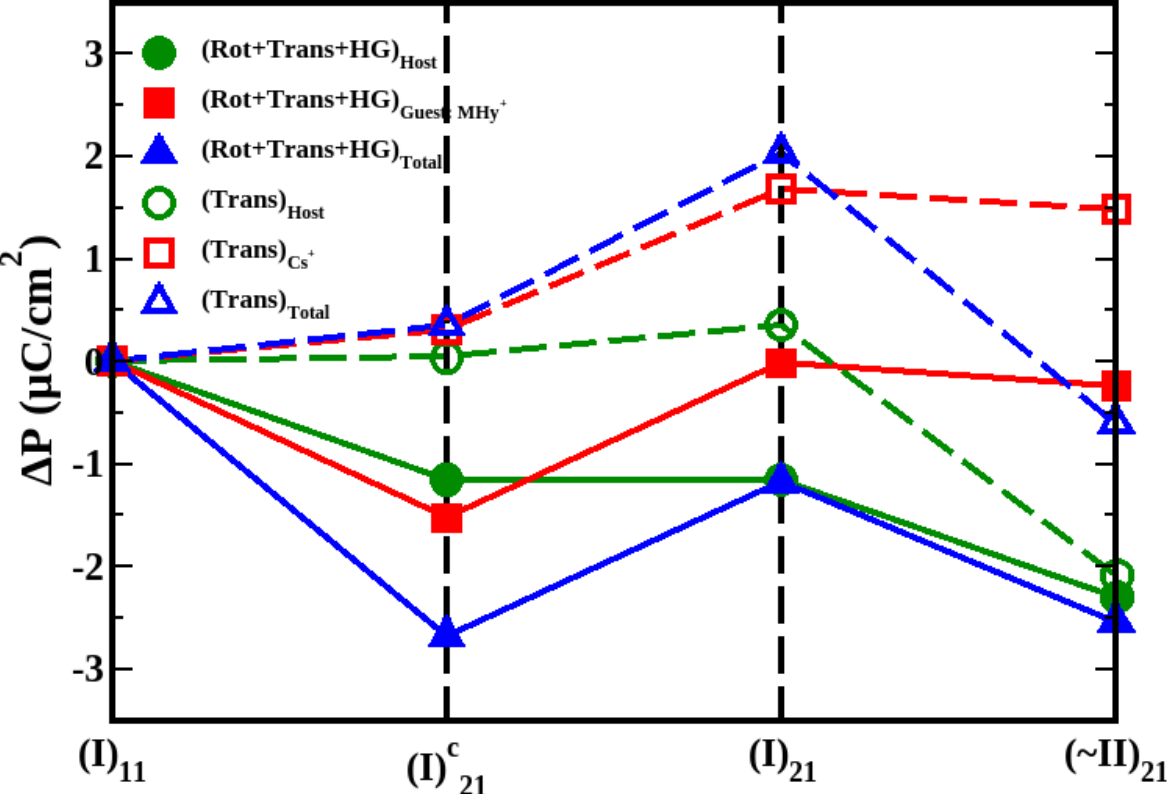}
	\caption{Net polarization in \ch{MHyPbCl3} across the path: $(I)_{11}\to(I)^c_{21}\to(I)_{21}\to(\sim II)_{21}$. Solid symbols represent total (blue), \textit{host} (green), and \textit{guest} (red) contributions to change in polarization in \ch{MHyPbCl3}. The open/hollow symbols represent the same contributions in \ch{CsPbCl3}. Organic \textit{guest} is substituted by an isotropic, rigid inorganic entity so as to separate out the translational component from the rotational and HG-coupling term that exists for an organic \textit{guest}. The lines are a guide to the eye.}
	\label{fig:delp_cs_mhy}
\end{figure}

\begin{figure}
	\centering
	\subfigure[]{
		\includegraphics[width=0.4\textwidth]{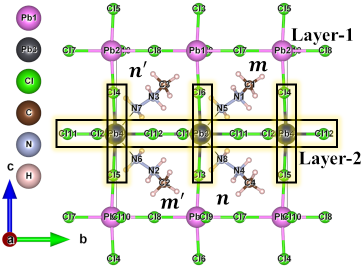}
		\label{subfig:atoms_more_dist_layer}
	}
	\hspace{0cm}
	\centering
	\subfigure[]{
		\includegraphics[width=0.4\textwidth]{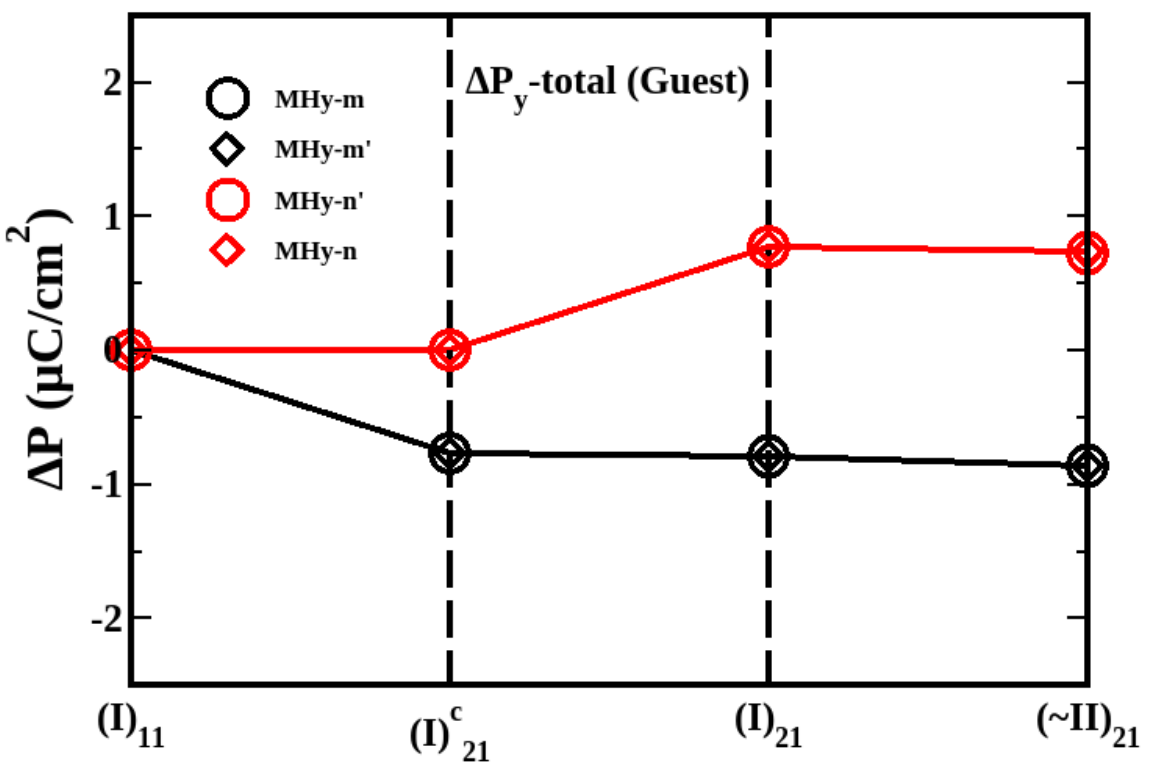}
		\label{subfig:py_mhy_tot}
	}
	\hspace{0cm}
	\centering
	\subfigure[]{
		\includegraphics[width=0.4\textwidth]{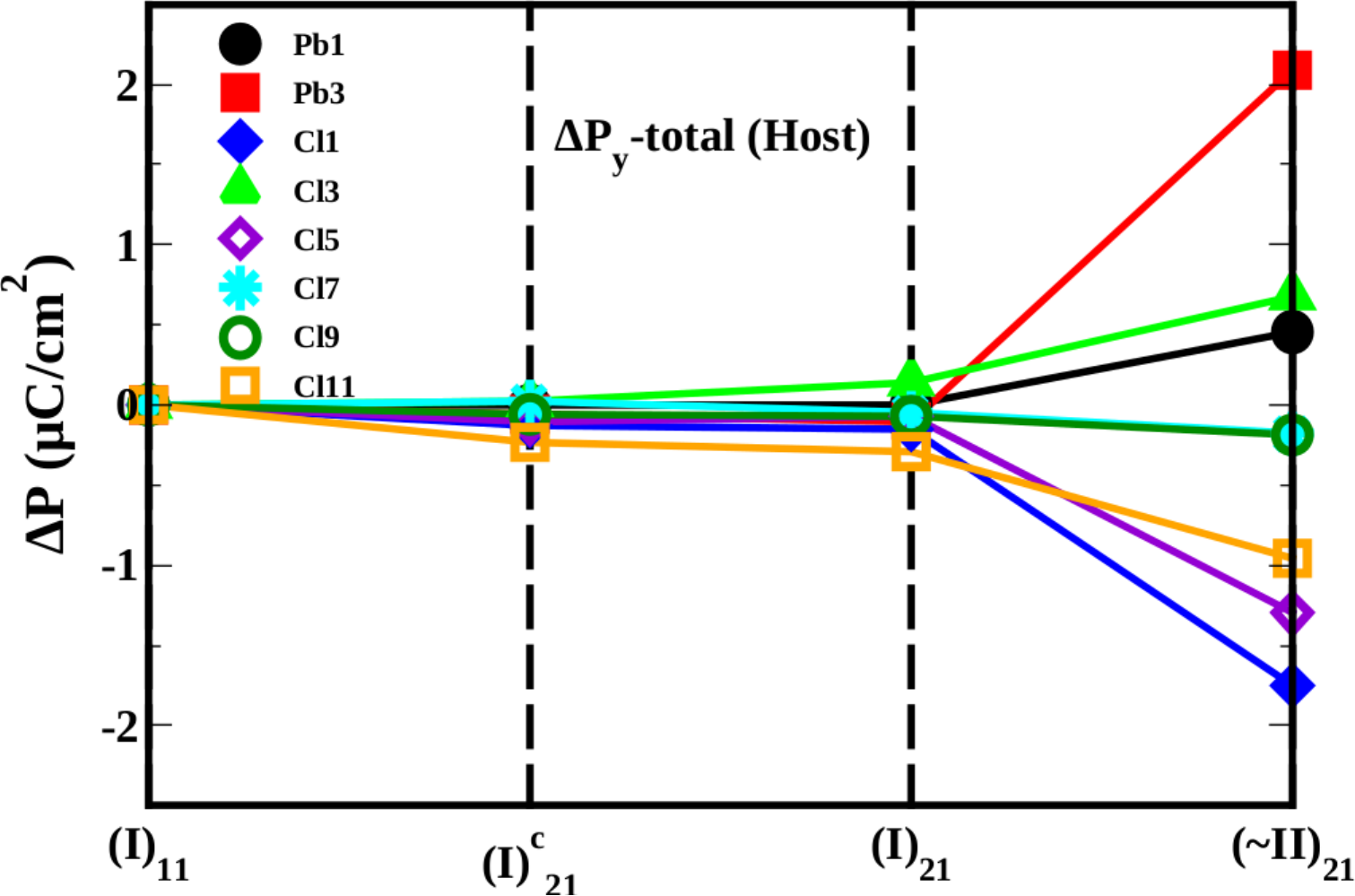}
		\label{subfig:py_lat_tot}
	}
	\caption{\label{fig:py_lat_mhy}~\subref{subfig:atoms_more_dist_layer} More distorted layer of the \textit{host} (Layer-2 as highlighted in the figure) contributes to polarization change most significantly. Quantification of the $y$-component of total polarization of~\subref{subfig:py_mhy_tot} \textit{Guest}.~\subref{subfig:py_lat_tot} \textit{Host}. Only $y$-component contributes to polarization in all the cases.}
\end{figure}

Various components of OIHPs can contribute to the polarization change, including \textit{guest}-reorientation, \textit{host}-distortion, and \textit{guest/host} coupling. The interplay of the \textit{guest} and \textit{host} of \ch{MHyPbCl3} cannot be ignored because of their synergistic roles in phase transitions due to the strong HG coupling. Hence, we use MLWFs to probe into the detailed mechanism regarding enhancement of polarization (For details, see Sec.~S-V B). Wannier centers (WCs) for the relevant structures are shown by magenta-colored spheres in Fig.~S24, which are used to extract the electronic polarization along with ionic contribution due to \textit{guest} ordering as well as \textit{host} configurations in different phases. Here, we have considered a path $(I)_{11}\to(I)^c_{21}\to(I)_{21}\to(\sim II)_{21}$ showing polarization change (Fig.~\ref{fig:delp_cs_mhy}) along the following sequence of events: (1) reorientation of the \textit{guest} at $m-m'$ in Phase-I, (2) \textit{guest} relaxation at $n-n'$ in Phase-I, and finally the (3) \textit{host} distortion involved in the phase transition. The Phase-I ($(I)_{11}$) has an absolute polarization of $8.96~\mu C/cm^2$ coming from both ionic and electronic components, which are all set to zero individually for referencing. In the final Phase-II ($(\sim II)_{21}$) structure, \textit{host} plays the dominating role in changing the polarization ($>90\%$), with only a nominal contribution from \textit{guest}. On the other hand, the unrelaxed intermediate structure ($I^c_{21}$) significantly reduces the polarization due to $m-m'$-reorientation ($\Delta P_{guest}=-1.54~\mu C/cm^2$), having slightly larger contributions than the octahedral \textit{host} ($\Delta P_{host}=-1.16~\mu C/cm^2$), as seen in Fig.~\ref{fig:delp_cs_mhy}. This might seem to agree with the experimental suggestion of \textit{guest} playing the major role in polarization enhancement~\cite{maczka2020three}. However, as seen in Fig.~\ref{subfig:py_mhy_tot}, further relaxation from the unrelaxed intermediate to relaxed-intermediate structure ($(I)^c_{21} \to (I)_{21}$) slightly affects the $m-m'$ contribution ($-0.04~\mu C/cm^2$), whereas changes the atomic positions at $n-n'$ significantly. The resultant polarization due to $n-n'$ ($1.54~\mu C/cm^2$) cancels the $m-m'$ contributions, thus effectively ruling out the role of the \textit{guest} in polarizing Phase-I. The same behavior is also seen in Phase-II (Fig.~\ref{subfig:py_mhy_tot} and Fig. S27). Thus, despite \textit{guest} being the primary order of the structural phase transition, it is the \textit{host} that plays a dominant role in polarizing the system. Our finding contrasts the suggestions from experiments~\cite{maczka2020three}, where the separation of roles of \textit{host/guest} in phase-transition and polarization was beyond the scope without the help of computational approach.
%
%

To segregate the effects of rotation and \textit{guest/host} coupling from translational contribution to polarization, each of the \ch{MHy+} ions are replaced by the inorganic atom, here, \ch{Cs+} at its center of mass. This also helps in filtering out the effect solely due to the translation of the dipole. 
Fig.~\ref{fig:delp_cs_mhy} demonstrates the exclusive role of rotation and HG coupling in drastically changing $\Delta P$ unlike translation ($1.48~\mu C/cm^2$) as in the inorganic halide perovskites. However, this translational contribution is suppressed due to its opposite nature of polarization with respect to contribution from \textit{guest}-reorientation and HG coupling. 
%
Next we look at the atomic contributions of the \textit{host} as it is responsible for $\sim 90\%$ total polarization change ($2.57~\mathrm{\mu C/cm}^2$) (Table~S15 and Fig.~\ref{subfig:atoms_more_dist_layer}, Fig.~\ref{subfig:py_lat_tot}, and Fig.~S30) which is quite significant in terms of the net polarization shown by similar OIHPs~\cite{stroppa2015ferroelectric,hu2017dipole}. 
We find that the collective effect of \ch{Pb} displacements and the polarization of \ch{Cl} electrons in the more distorted layer of the \textit{host} (layer-2) (Fig.~\ref{subfig:py_lat_tot}) contributes the most towards enhancement of polarization as the structure transitions from $(\sim II)_{21}\to (I)_{11}$. Furthermore, we also find that out of the $s$ and the $d$ non-bonding electrons of \ch{Pb}, the former plays the most significant role. This dominance of $s$ lone pair was also highlighted recently in~\cite{fabini2020underappreciated,huang2022understanding} by showing that the lone-pairs around \ch{Pb}-atoms are highly directional in nature.  



\subsubsection{Polarization feedback (Effect of the \textit{host} on the \textit{guest})}
So far, having investigated the effect induced on the \textit{host} due to \textit{guest} reorientation in terms of both structural and polar properties, we next ask if the \textit{host} could further induce polarization in the \textit{guest} due to phase transition from $(I)_{11} \to (\sim II)_{21}$. This effect is bifurcated into two primary contributions involving (1) \textit{host} introduction and (2) \textit{host} distortion. 

To understand the former effect, we first consider only the \textit{guest}-reorientation ($(I)_{11}\to(I)^c_{21}$, \textit{i.e.}, $0^{th}$ and $10^{th}$ images in Fig.~\ref{fig:pol_mhy_berry_gaussian}), where total polarization of isolated \textit{guest}s (dipoles) and the combined \textit{guest/host} system are compared. This comparison at the final structure ($10^{th}$ image) results in a polarization difference of $1.16~\mu C/cm^2$ because of \textit{host} introduction. This is solely due to the electronic component as the ionic contributions are canceled through the same \textit{guest} and \textit{host} coordinates (see SI Sec.~S-V C for details). This difference consists of two factors. The first one is the pure \textit{host} electronic polarization as a primary effect, whereas the secondary effect of the \textit{host} on \textit{guest} electronic polarization is termed feedback polarization. The latter is computed by comparing the \textit{guest} electronic polarization of the $10^{th}$ images in Fig.~\ref{fig:pol_mhy_berry_gaussian}, which amounts to only $0.02~\mu C/cm^2$ (roughly $1\%$ of the total polarization). 

\begin{figure}
	\centering
	\includegraphics{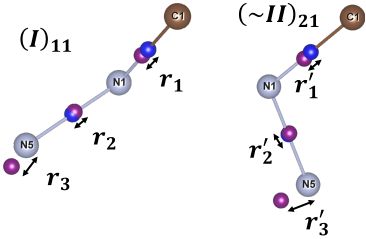}
	\caption{Schematic representing shift in bond- and atom-centered WCs between \textit{guest} ions of two structures. Here, we have taken one of the structures as $(I)_{11}$ for clearly demonstrating the shift in WCs between two structures. In actual, we have considered $(I)_{21}$, and $(\sim II)_{21}$ to calculate $\delta (\Delta WC)$. The violet and blue spheres represent the WCs and bond centers, respectively.  $r_1,~r_2,~r'_{1},~r'_{2}$ are the distances with respect to the B.C., $r_3$ and $r'_{3}$ are the WC shift with respect to A.C.}
	\label{fig:ht_12_wc_shift_pict}
\end{figure}

The second part of the feedback effect occurs when the \textit{host} undergoes distortion. This is computed for the identical \textit{guest} coordinates, which would eliminate any contributions from the ionic part of the \textit{guest} polarization. The resultant change in the \textit{guest} polarization, therefore, will only be due to the changes in the electronic cloud on the \textit{guest} caused by \textit{host} distortion. We achieve this by comparing the WCs associated with the \textit{guest} in the structures, $(I)_{21}$ and $(\sim II)_{21}$. For this, we have first assigned the WCs to each atom, depending on whether they are bond-centered (B.C.) or atom-centered (A.C.). 
The WC type can be decided by the ratio of the distance of the WC from the two atoms it is in the closest proximity with. We suggest that, for \ch{MHyPbCl3}, an atom-centered WC would have a ratio $< 0.25$, while higher values of this ratio would imply a bond-centered WC. The shift in the WCs (Fig.~\ref{fig:ht_12_wc_shift_pict}) in the two structures would, hence, directly contribute to feedback polarization due to \textit{host} distortion, defined as: 
\begin{equation}
\delta (\Delta WC)=\left[(\Delta WC)_{(\sim II)_{21}}-(\Delta WC)_{(I)_{21}}\right]
\label{eq:del_delWC}
\end{equation}
where $(\Delta WC)_{(\sim II)_{21}}=(WC)_{(\sim II)_{21}}-\mathrm{B.C./A.C.}$ for the $(\sim II)_{21}$-structure.

We find that this feedback amounts to $-0.22~\mu C/cm^2$ (Fig.~S32 and Table~S17, exists only along $y$-direction, cancels out in other directions), which demonstrates a further reduction of net polarization in $(\sim II)_{21}$ ($\sim 9\%$ with respect to the total polarization). 
Since \textit{host} distortion in itself is responsible for bringing down the polarization (as shown earlier), followed by a further reduction due to feedback effect, therefore the latter acts as positive feedback. This effect is significant given the relatively small polarization offered by OIHPs in general, which is distributed on the \textit{guest} at $m-m'$ and $n-n'$ in a ratio of 1:0.2. 

\subsection{\label{sec:dos_bands}Coupling of polar and electronic properties}

After probing the polar properties of \ch{MHyPbCl3} and determining the elements that are critically involved in changing the polarization as the phase transition occurs, we now look at the correspondence between the polar and the electronic properties in terms of both \textit{guest}, and \textit{host}. (See SI Sec.~S-VI for details related to band gaps).

First, we focus on the contribution from \textit{host} atoms to electronic states. A one-to-one correspondence between polar and electronic properties is expected as the \textit{host} dominates total polarization. Indeed, 
the \textit{host} shows similar significant contributions in electronic structure. This is evident from PDOS (partial density of states), where the $p$-orbitals of \ch{Cl} and \ch{Pb} of the more distorted layer, $i.e.$, layer-2 (also responsible for polarization) affect the valence band significantly (Fig.~S34 and Fig.~S35)~\cite{fan2015perovskites}. 

\begin{figure*}
	\centering
	\subfigure[$(I)_{21}\to(\sim II)_{21}$]{
		\includegraphics[width=0.42\textwidth]{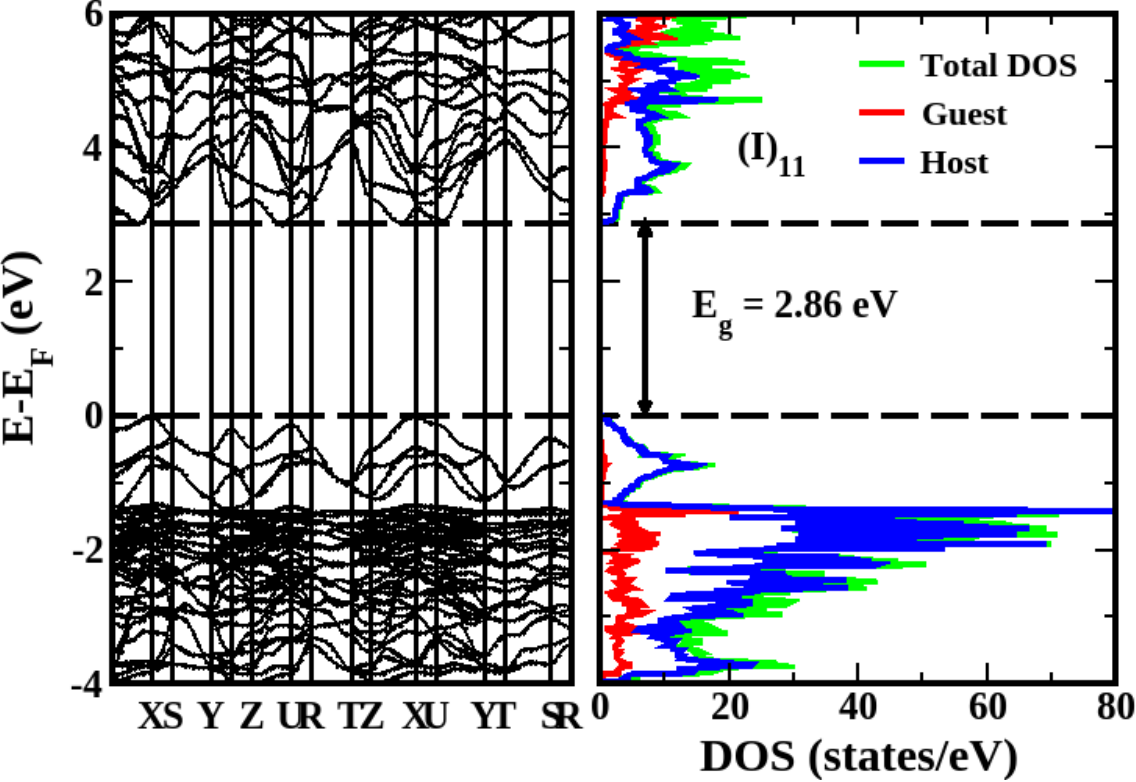}
		\label{subfig:h2g2_mhy_feedback_pdos}
	}
	\hspace{1cm}
	\centering
	\subfigure[$(I)_{11}\to(\sim II)_{11}$]{
		\includegraphics[width=0.42\textwidth]{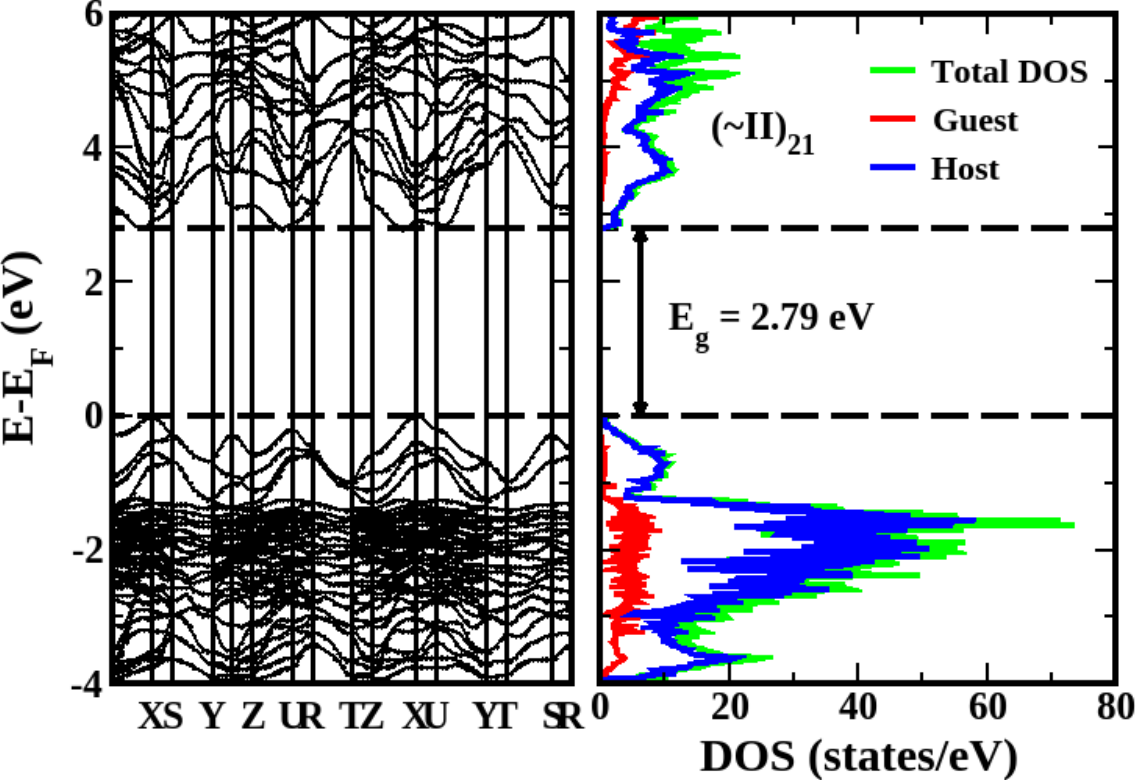}
		\label{subfig:h2g1_mhy_feedback_pdos}
	}
	\caption{\label{fig:feedback_mhy_pdos}Effect of polarization feedback (due to \textit{host} distortion) on the energy levels of \textit{guest} (\ch{MHy}-ions) near the band-edges for orientation of the \textit{guest} in both the phases, Phase-I and Phase-II.}
\end{figure*}

Secondly, we explicitly look at the effect on the \textit{guest} density of states due to \textit{host} distortion in the form of feedback polarization. This effect is rather unusual, as in general, the \textit{host} atoms populate the band-edges in other OIHPs, whereas the organic cations have almost an insignificant contribution~\cite{mehdizadeh2019role,kim2020first}. However, here, we show that this large \textit{guest}, \ch{MHy+} (Fig.~\ref{fig:feedback_mhy_pdos}) directly affects the electronic properties through the polarization feedback by the \textit{host}. Fig.~\ref{fig:feedback_mhy_pdos} shows that the participation of the \textit{guest} atoms near the band edges gets affected substantially depending on the orientation of the molecule. For \textit{guest} orientation similar to Phase-II and Phase-I, \textit{guest} DOS increases with the decrease in feedback polarization in both cases (Fig.~\ref{subfig:h2g2_mhy_feedback_pdos} and Fig.~\ref{subfig:h2g1_mhy_feedback_pdos}, respectively). Although in the former, the feedback correlates positively with \textit{host} distortion, whereas in the latter, this correlation is negative. Thus, the effect of the \textit{guest} on the electronic structure can be tuned by controlling feedback through \textit{host} distortion in both phases of \ch{MHyPbCl3}.
This feedback on the \textit{guest} at the band edges has direct effects on quasiparticle band gaps and exciton binding energies of 3D hybrid perovskites. Such direct roles are recently discovered in 2D perovskites but the mechanism through feedback is not yet explored~\cite{mahal2023energy,gan2023role,fish2023impact,filip2022screening}. The orientation and packing of the \textit{guest} also guide the \textit{host} configuration as well as the extent of metal-halide orbital overlap, thus also affecting the electronic structure indirectly. A recent study showed the influence of the \textit{guest} on material response, such as phase transitions, photoluminescence emission, etc.~\cite{cuthriell2023cyclic}. The compositions with linear spacers undergo distinct phase transitions, whereas cyclic spacers (having large steric hindrances) lack first-order transition for the investigated temperature range. Additionally, both of them show similar photo-luminescence properties. Therefore, modulation of the spacer/\textit{guest} ion can lead to a larger structural and thermal phase space availability, thereby enhancing the functionalization and applicability of OIHPs.

\subsection{\label{sec:conclusion} Conclusions}
This work investigates the driving force behind the structural phase transition and the origin of polarization enhancement with increase in temperature in the super-tolerant and strongly coupled member of hybrid halide perovskite, \ch{MHyPbCl3}, from the perspective of \textit{host/guest} system.
The \textit{guest} reorientation, quantified by comparing barriers between the two phases, $(I)_{11}$ and $(\sim II)_{21}$, emerges as the primary order parameter. The substantial rotational barrier is also an indication of slow and sticky \textit{guest} dynamics. A transition character, defined as $\lambda$, quantifies the \textit{guest/host} contribution, highlighting the \textit{guest} dominance. The \textit{host} atoms are primarily responsible for polarization enhancement, with the \textit{guest} contributing only minimally. The \textit{host} atoms of the more distorted layer of octahedra are found to be considerably involved, thereby establishing a one-to-one correlation between structural and polar properties. An interesting positive feedback effect that the \textit{host} induces on the \textit{guest} polarization is observed. This would mean that as the system transitions from a more to a less distorted state ($(\sim II)_{21}\to (I)_{21}$), there is an enhancement of polarization due to \textit{guest}-reorientation through the \textit{host} distortion. As a consequence of this, the \textit{host} further induces a polarization on \textit{guest}, leading to a secondary enhancement.  This feedback is also coupled with the energy levels of the \textit{guest} at the band edges and is dependent on the starting \textit{guest} orientation. This suggests that {\it guest} tunability can directly influence electronic and optical properties in over-tolerant perovskites. Hence, these systems offer viable alternatives for improved photovoltaic applications.

\section{\label{spsec:comp_details}Computational Details}

Self-consistent total-energy calculations were carried out for different structures using ultrasoft pseudopotential~\cite{rappe1991optimised}. The lattice parameters were compared with the experimental values using two Generalized Gradient Approximation (GGA) exchange-correlation energy functionals, PBE~\cite{perdew1996generalized} and PBE functional revised for solids (PBEsol)~\cite{perdew2008restoring} as implemented in Quantum-ESPRESSO (QE) computational package~\cite{giannozzi2009quantum,giannozzi2017advanced}. The corrections due to dispersion interactions in density functional theory (DFT) were employed through the second-generation (D2) van der Waals (vdW) corrections proposed by Grimme~\cite{grimme2006semiempirical,barone2009role}.  

For ultrasoft pseudopotentials, the electronic wave functions were expanded in plane waves upto a charge density cut-off of $480~Ry$ and a kinetic energy cut-off of $80~Ry$. A cut-off of $4\times2\times2$ Monkhorst-Pack $k$-point mesh was chosen for Brillouin zone integration depending on the ordering of the lattice parameters. The structures were relaxed until the Hellmann-Feynmann forces were less than $0.026~meV/\text{\AA}$. Polarization calculations were done using the Berry phase approach with $10$ k-points along each direction~\cite{king1993theory,resta1994modern,resta2007theory,spaldin2012beginner}. Phase-I is taken as the reference for the polarization of all structures. Maximally localized Wannier functions were obtained to evaluate polarization using WANNIER90 code~\cite{marzari1997maximally,souza2001maximally,pizzi2020wannier90} integrated with QE. Wannier functions were generated using only valence bands, with a convergence criterion of $10^{-10} \text{\AA}^2$. The starting projections included $s,~p$ and $d$ orbitals for \ch{Pb}, $p$ for \ch{Cl}, $sp^3-$ hybridization for \ch{C}, $sp^3$ for each of terminal \ch{N} atoms and $s-$ orbital for the two H-atoms bonded with middle nitrogen of each \ch{MHy}-ion. Hybrid-PBE0 functional~\cite{perdew1996rationale,yang2018hybrid} was also used to verify the lattice parameters and the difference in energies between the two phases, all implemented in VASP~\cite{kresse1993ab,kresse1996efficiency,kresse1996efficient} using the projector augmented wave (PAW) pseudopotential~\cite{kresse1994norm}. For PAW-pseudopotentials, kinetic energy cut-off $750~\mathrm{eV}~(\approx55~\mathrm{Ry})$ and k-mesh cut-off same as that of ultrasoft pseudopotential was chosen. The phonon dispersion curves for the structures were obtained using \textit{ab initio} lattice dynamics with the finite displacement method on a $2\times2\times2$ supercell with PHONOPY package~\cite{phonopy}. On the optimized \textit{guest} coordinates for different structures, we used Gaussian 16 program package~\cite{g16} implementing density functional three-parameters hybrid, B3LYP methods~\cite{becke1996density,lee1988development} with the 6-311G(2d,2p)~\cite{mclean1980contracted,krishnan1980self,ciupa2020vibrational} basis set for evaluating polarization solely due to the \textit{guest}. Climbing Image-Nudged Elastic Band (CI-NEB) has been performed using VASP to determine the path of phase transition and the corresponding energy barrier.

\begin{acknowledgements}
The authors gratefully acknowledge computational resources provided by IISER Bhopal as well as “PARAM Shivay” at Indian Institute of Technology (BHU), Varanasi, which is implemented by C-DAC and supported by the Ministry of Electronics and Information Technology (MeitY) and the Department of Science and Technology (DST), Government of India, through National Supercomputing Mission (NSM). P.S. acknowledges funding through UGC-JRF (India) for carrying out the Ph.D. program. S.M. acknowledges funding through the Integrated Ph.D. program at IISER Bhopal. 
\end{acknowledgements}
\bibliographystyle{unsrt}
\bibliography{aps_bib}

\end{document}